\documentclass{aa}
\usepackage{graphicx}
\usepackage{txfonts}
\usepackage{natbib}

\begin{document}

\title{APEX CO(3-2) observations of NGC6822}

\author{De Rijcke S. \inst{1}\fnmsep\thanks{Postdoctoral Fellow of the
Fund for Scientific Research - Flanders (Belgium)(F.W.O)} \and Buyle
P. \inst{1} \and Cannon J. \inst{2} \and Walter F.  \inst{2} \and
Lundgren A.  \inst{3} \and Michielsen D. \inst{4} \and Dejonghe
H.\inst{1} }

\offprints{Sven De Rijcke, Pieter Buyle}

\institute{Astronomical Observatory, Ghent University, Krijgslaan 289
  - S9, B-9000 Ghent, Belgium \and Max-Planck-Institut f\"ur
  Astronomie, K\"onigstuhl 17, D-69117 Heidelberg, Germany \and ESO,
  Casilla 19001, Santiago 19, Chile \and University of Nottingham,
  University Park, Nottingham, NG7 2RD, UK \\
  \email{Sven.DeRijcke@UGent.be,Pieter.Buyle@UGent.be}}

   \date{Received September 15, 1996; accepted March 16, 1997}

   \abstract{We observed the $^{12}$CO(3$\rightarrow$2) emission of
the emission-line regions Hubble\,I, Hubble\,V, Hubble\,X,
Holmberg~18, and the stellar emission-line object S28 in NGC6822 with
the ESO Atacama Pathfinder Experiment (APEX) 12m telescope as part of
its science verification. The very low system temperature of
$130-180$~K enabled us to achieve detections in 4 single pointings and
in a high spatial resolution $70''\times 70''$ map of Hubble\, V. We
compare the spectra with H{\sc i} observations, obtained with the
Australia Telescope Compact Array, of the same regions. In combination
with previous multi-line CO observations, we perform a preliminary
investigation of the physical conditions in Hubble\, V using a simple
LTE model. We estimate the mass of the Hubble~V region and the
H$_2/I_{\rm CO(3\rightarrow 2)}$ conversion factor. Also, we show that
Hubble~V is located very near the line-width versus size relation
traced by the Milky Way and LMC molecular clouds.

   }

   \maketitle
%

\section{Introduction}

We present $^{12}$CO(3$\rightarrow$2) observations of the
emission-line regions Hubble\,I, Hubble\,V, Hubble\,X, Holmberg~18,
and the stellar emission-line object S28 \citep{kd} in NGC6822 with
the ESO Atacama Pathfinder Experiment (APEX) 12m
telescope\footnote{This publication is based on data acquired with the
Atacama Pathfinder Experiment (APEX). APEX is a collaboration between
the Max-Planck-Institut f\"ur Radioastronomie, the European Southern
Observatory, and the Onsala Space Observatory.}. NGC6822 is a Local
Group dwarf irregular galaxy, of type IB(s)m. Its study started with
the landmark paper of \cite{h25}, who identified it as a galaxy in its
own right, external to the Milky Way. The most recent estimate of its
distance, based on observations of 116 Cepheid variables, places it at
$466 \pm 20$~kpc \citep{p04}. \cite{ppr05} find a metallicity $12 +
\log($O/H$) = 8.37 \pm 0.09$ for Hubble V and $12 + \log($O/H$) = 8.19
\pm 0.16$ for Hubble X, corresponding to roughly half the solar
metallicity. Star formation proceeded at an almost constant rate up to
the present, except for the central bar region, where star-formation
increased by a factor of $3-4$ during the last 600~Myr
\citep{w01}. Its proximity allows us to study the
different components and phases of its interstellar medium on scales
of order 10$-$100 parsec.

The detection of the compact molecular clouds associated with
Hubble\,V was first reported by \cite{w94}. Later on, the
emission-line regions Hubble\,I, Hubble\,V, and Holmberg~18, and the
stellar emission-line object S28 were observed in
$^{12}$CO(1$\rightarrow$0) emission by \cite{i97}. Moreover, for
Hubble\,V, the brightest HII region in NGC6822, detections of
$^{12}$CO(2$\rightarrow$1), $^{12}$CO(3$\rightarrow$2),
$^{12}$CO(4$\rightarrow$3), and $^{13}$CO(1$\rightarrow$0) have been
reported \citep{i03}. $^{12}$CO(1$\rightarrow$0) and
$^{12}$CO(2$\rightarrow$1) observations centered on Hubble\,X, on the
other hand, did not yield a detection \citep{i03}. These results will
serve as a comparison for the APEX data presented here.

In this Letter, we combine the $^{12}$CO(3$\rightarrow$2) line
intensities measured with APEX with line intensities of other
$^{12}$CO and $^{13}$CO transitions, taken from the literature, to
constrain the physical conditions of the molecular interstellar medium
of NGC6822 using simple LTE models. We also investigate the spatial
distribution of the $^{12}$CO(3$\rightarrow$2) emission and how it
correlates with previous high resolution HI observations.

\begin{figure*}
\vspace*{9cm}
\special{hscale=35 vscale=35 hsize=540 vsize=560
hoffset=-20 voffset=45 angle=0 psfile="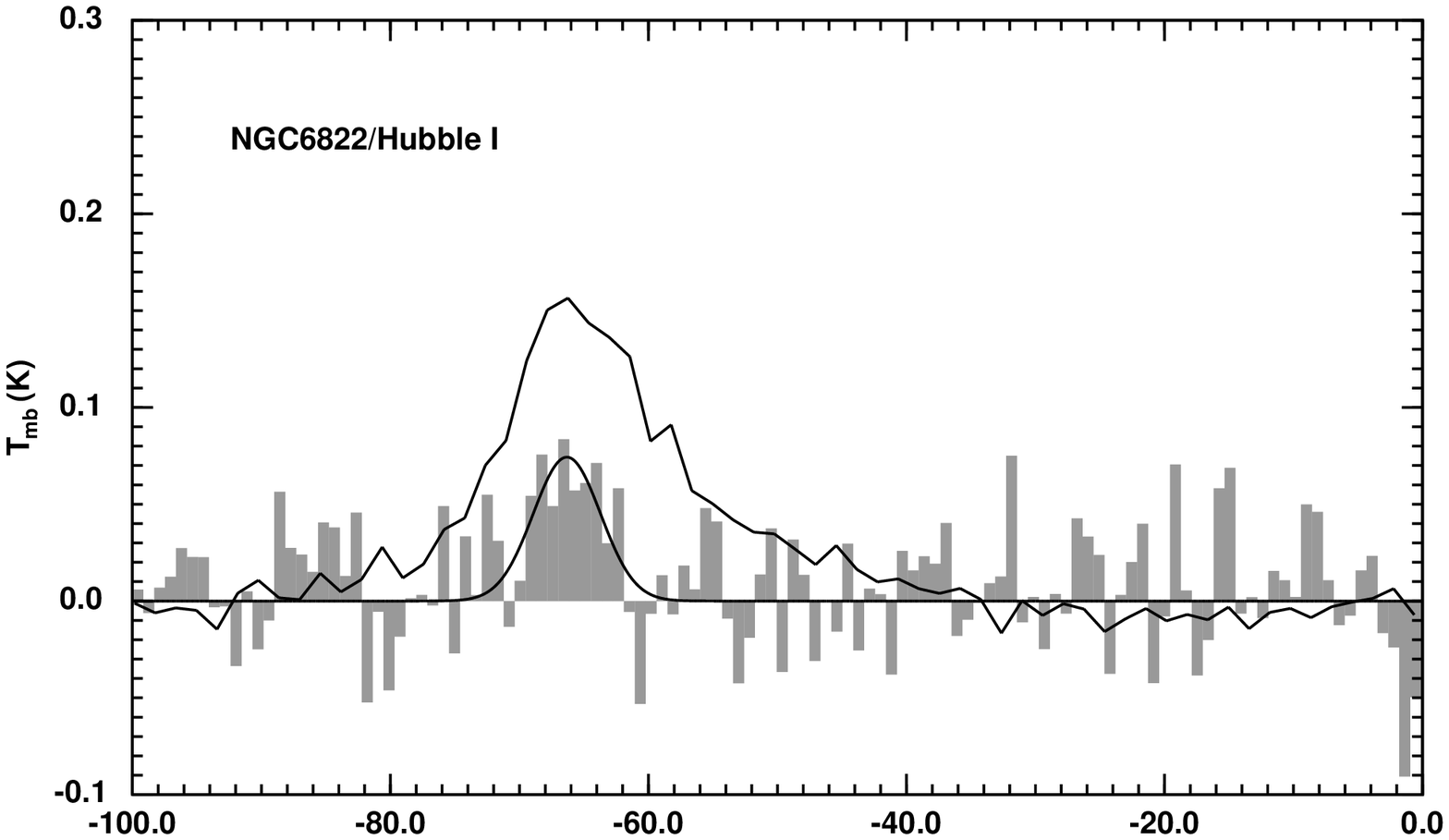"}
\special{hscale=35 vscale=35 hsize=540 vsize=560
hoffset=235 voffset=45 angle=0 psfile="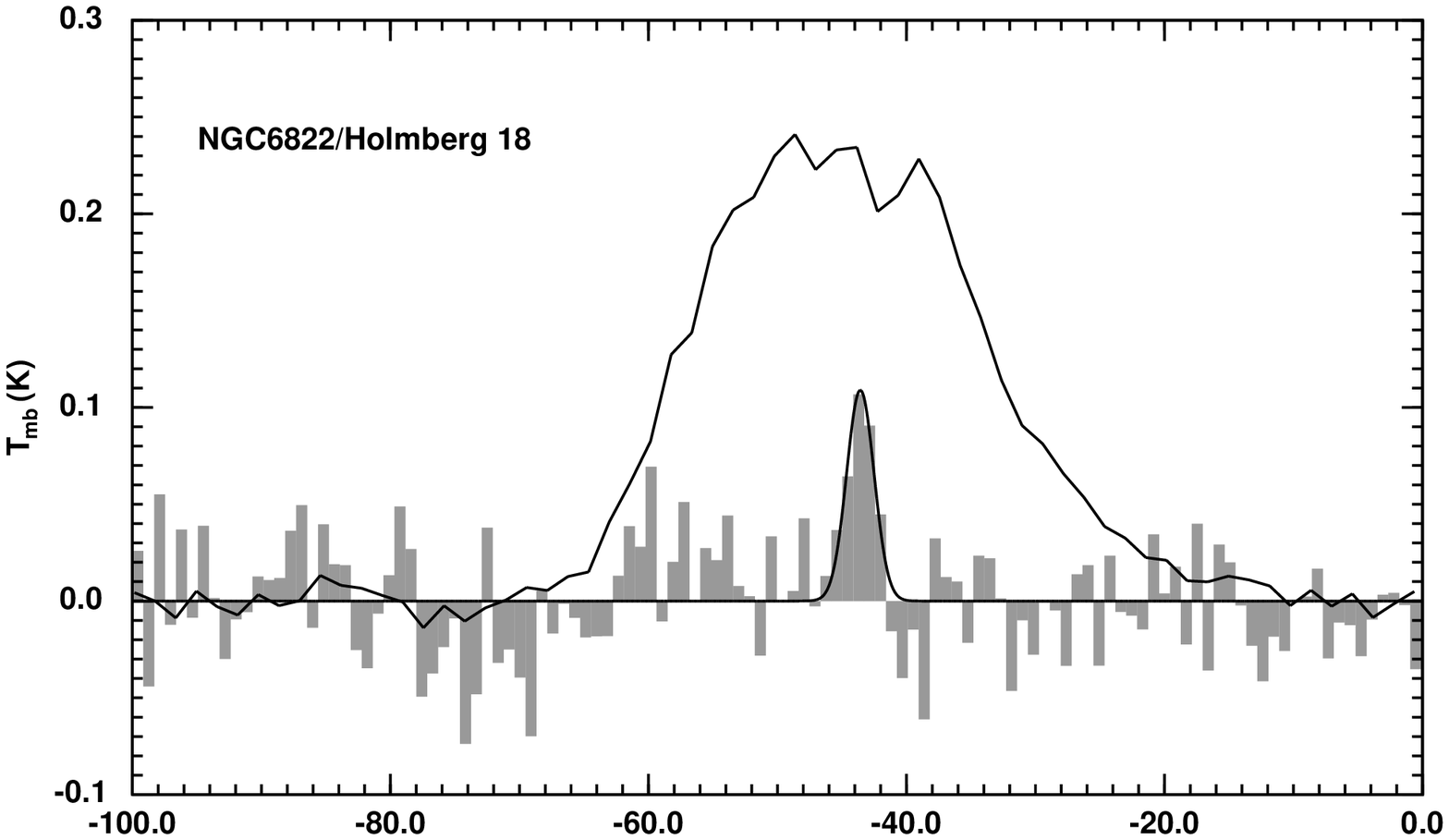"}
\special{hscale=35 vscale=35 hsize=540 vsize=520
hoffset=-20 voffset=-70 angle=0 psfile="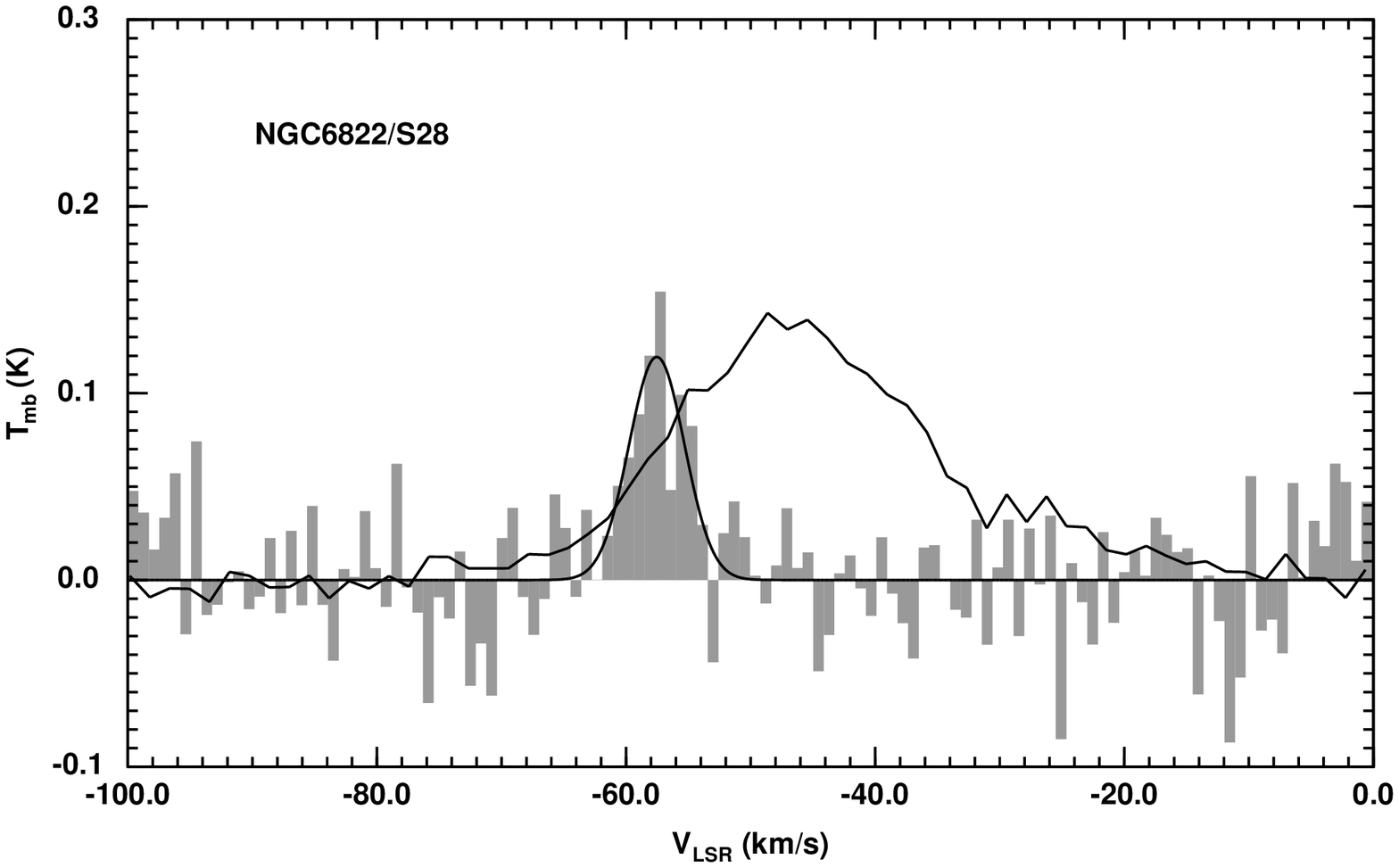"}
\special{hscale=35 vscale=35 hsize=540 vsize=720
hoffset=235 voffset=-70 angle=0 psfile="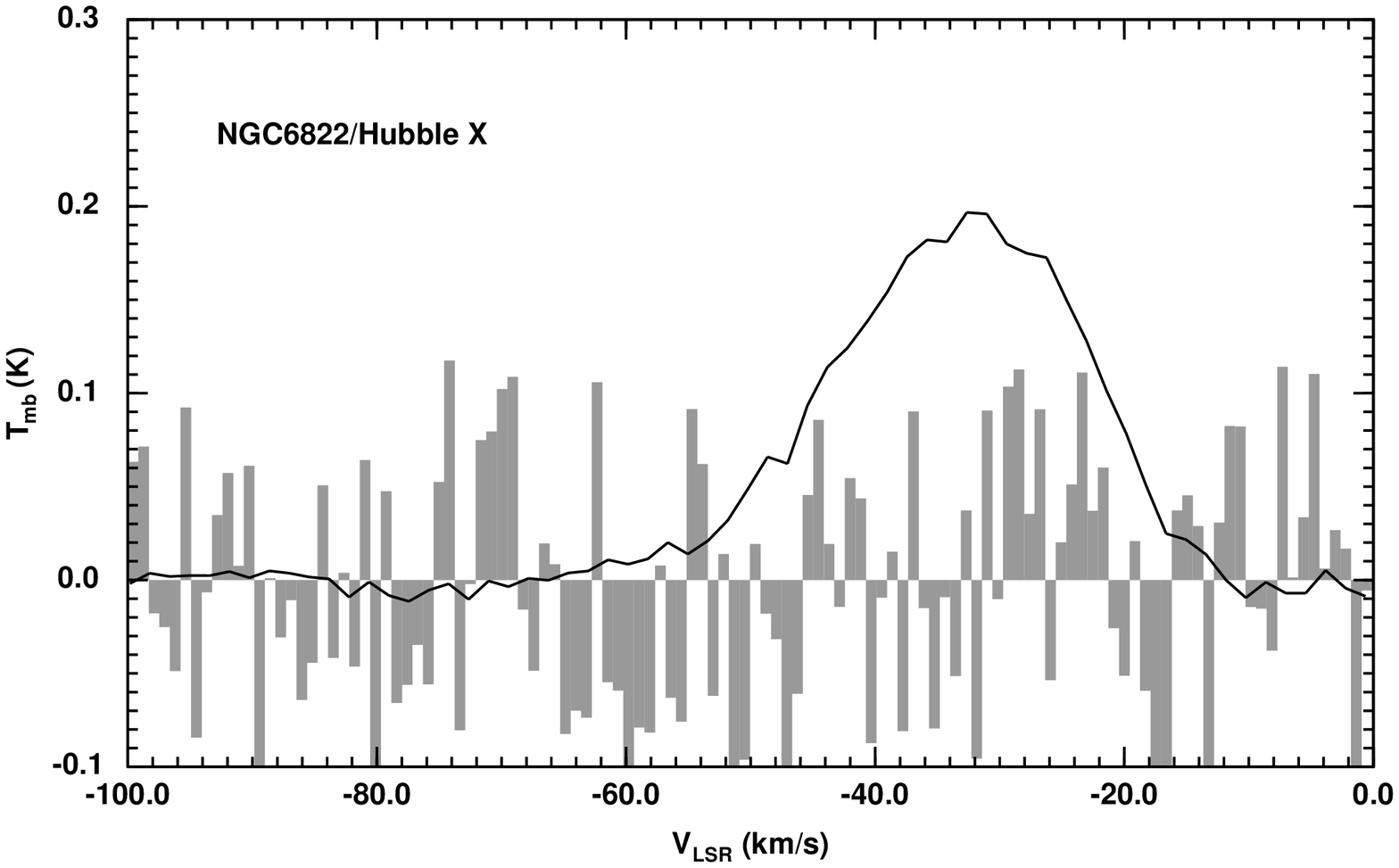"}
\caption{CO(3$\rightarrow$2) spectra of the star-forming regions
Hubble\rm{I}, Hubble\rm{X}, and Holmberg~18 and the stellar
emission-line object S28 in NGC6822, overplotted with the best fitting
Gaussian. The black curve indicates the H{\sc i} emission as derived
by \cite{deblok06}. Evidently, the molecular gas associated with
Hubble~I and Holmberg~18 has the same velocity as the neutral gas. The
velocity of the molecular gas associated with the stellar object S28
differs from that of the H{\sc i}. Hubble~X was not detected.
\label{HIV}}
\end{figure*}

\section{Observations and data reduction} 

We observed the $^{12}$CO(3$\rightarrow$2) line towards the star
forming regions Hubble\, \rm{I}, \rm{V}, \rm{X}, and Holmberg~18, and
the stellar emission-line object S28 \citep{kd} in NGC6822 with the
ESO Atacama Pathfinder Experiment (APEX) 12m telescope on the nights
of 17, 18, 20, 21, 25 and 26 August 2005 as part of its science
verification. We used a Fast Fourier Transform Spectrometer (FFTS)
backend built by MPIfR \citep{k06}. The frequency was centered on
CO(3$\rightarrow$2) (345.79\, GHz) and corrected for the systemic
velocity of NGC6822. The beamsize of the telescope at this frequency
is 18.2$''$ FWHM \citep{g06}. We used a 1~GHz bandwidth divided over
8192 channels, corresponding to a velocity resolution of 0.1~km\
s$^{-1}$ per channel. All targets were observed with a single
pointing, except for Hubble\rm{V}, which was spatially mapped on a
rectangular Nyquist $1'\times1'$ grid. On-source integration times
were 31.4~minutes (Hubble~I), 9.1~min (single pointing of Hubble{\rm
V}), 1.0-10.4~min per pointing of the map of Hubble~V, or 110.1~min in
total, 3.1~min (Hubble{\rm X}), 21.9~min (Holmberg 18), and 22.3~min
(S28). The rms of the pointing model was about 3$''$. Pointing was
regularly checked and updated on the nearby pointing source W-Aql. We
used position switching. The reference position was set to +480$''$ in
R.A. relative to the center position of each source. Dual side band
(DSB) system temperatures were 128~K (Hubble{\rm I}), 152~K (single
pointing of Hubble{\rm V}), 135-182~K (map of Hubble{\rm V}), 133~K
(Hubble{\rm X}), 156~K (Holmberg 18), and 176~K (S28) \citep{r06}. The
precipitable water vapour was 0.5-1.0~mm, corresponding to a
$\tau_{225}$ of 0.27-0.29, during the observations. Calibration errors
amount to $\sim 15$\%. In order to enhance the signal-to-noise ratio
without compromising the spectral resolution, the spectra were
rebinned to a 0.8~km\ s$^{-1}$ resolution and they have been rectified
over the velocity range [$-100,0$]~km~s$^{-1}$.


The data reduction was performed with the standard data analysis
program GILDAS of the 30m IRAM radio telescope. Observations related
to the same pointing were first added together. Afterwards a
polynomial of 4th order was fitted to an emission line free region of
the spectral baseline and subtracted off this baseline. Some spectra
still showed a residual double sinusoidal variation. One of these
ripples arose due to a vibration of the gore-tex membrane that covers
the entrance window to the Cassegrain cabin (L. Nyman, priv. comm.)
and the other due to the vibration of the cold head of the
closed-cycle cooling machine, which affected the LO coupling of the
receiver, causing the receiver gain to vary \citep{r06}. We fitted a
combination of two sine functions to those spectral baselines that
contained this sinusoidal variation, omitting the spectral region
around the CO(3$\rightarrow$2) emission line, and subtracted this off
the spectrum. The antenna temperatures, $T_A^*$, were converted to to
main-beam brightness temperatures ($T_{\rm mb}=T_A^*/\eta_{\rm mb}$),
using the main-beam efficiency $\eta_{\rm mb} = 0.7$.

\section{Discussion}

\begin{table*}
\caption{CO(3$\rightarrow$2) properties of the targeted regions in
NGC6822:~the peak main-beam temperature, $T_{\rm mb}$, the velocity of
the line with respect to the Local Standard of Rest, the line FWHM,
and the integrated line intensity, $I_{\rm CO}$ .
\label{tabres}}
\begin{center}
\begin{tabular}{|c|cccccc|} \hline
name & RA & DEC & $T_{\rm mb}$ (K) & LSR velocity (km/s) & FWHM (km/s)
& $I_{\rm CO}$ (K~km~s$^{-1}$) \\ \cline{1-7} Hubble\rm{I} & 19 44
31.64 & -14 42 01.2& $0.07 \pm 0.02$ &$-66.3 \pm 0.4$ & $6.1 \pm 1.1$
& $0.49 \pm 0.11$ \\ Hubble\rm{V} & 19 44 52.80 & -14 43 11.0 & $0.61
\pm 0.02$ &$-41.3 \pm 0.1$ & $6.0 \pm 0.2$ & $3.89 \pm 0.14$ \\
Hubble\rm{V} & 19 44 52.80 & -14 43 11.0 & deconvolved &$-41.3 \pm
0.4$ & $6.0 \pm 0.2$ & $6.65 \pm 0.59$ \\ Hubble\rm{X} & 19 45 05.20 &
-14 43 13.0 & & & & $< 0.30$ \\ Holmberg~18 & 19 44 48.93 & -14 52
38.0 & $0.11 \pm 0.02$ &$-43.6 \pm 0.2$ & $2.5 \pm 0.4$ & $0.29 \pm
0.07$ \\ KD82\_S28 & 19 44 57.79 & -14 47 51.5 & $0.12 \pm 0.02$
&$-57.5 \pm 0.3$ & $5.2 \pm 0.7$ & $0.67 \pm 0.11$ \\ \cline{1-7}
\end{tabular}
\end{center}
\end{table*}

We fitted Gaussians to the detected emission lines in order to
estimate the peak intensity, $T_{\rm mb}$ (K), and the integrated
intensity, $I_{\rm CO} = \int T_{\rm mb}(v)\,dv$ (K~km~s$^{-1}$), of
the $^{12}$CO(3$\rightarrow$2) emission lines of the observed
star-forming regions (see Table \ref{tabres}). We used the best
fitting Gaussian and the 1$\sigma$ noise on the spectrum, estimated
from the spectral region between $-100$ and 0~km/s and excluding the
emission line, to generate 1000 new noisy spectra. These were analysed
the same way as the original spectrum, allowing us to estimate the
1$\sigma$ errors on these quantities. For the non-detected
star-forming region Hubble\rm{X}, we give a 3$\sigma$ upper limit over
a velocity width of 6~km/s.

\subsection{Physical conditions in Hubble{\rm V}}

\begin{figure}
\vspace*{8cm}
\special{hscale=50 vscale=50 hsize=540 vsize=520
hoffset=-25 voffset=-37 angle=0 psfile="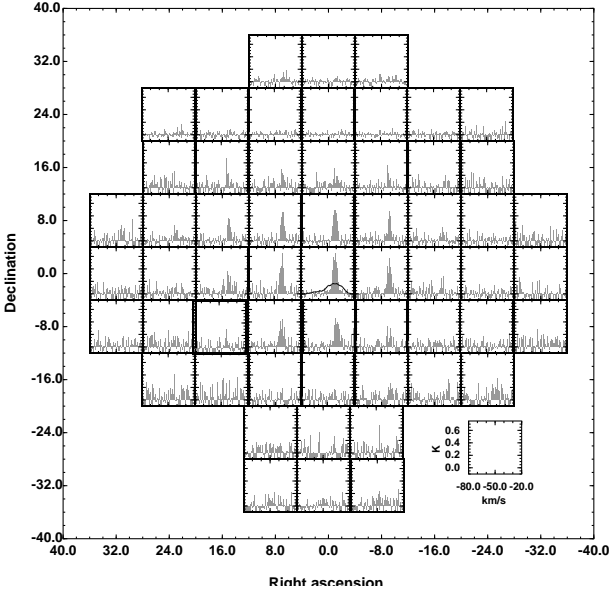"}
\caption{$^{12}{\rm CO}(3\rightarrow2)$ map of Hubble\rm{V}. Each
panel shows the brightness temperature $T_{\rm mb}$ as a function of
velocity with respect to the Local Standard of Rest (LSR). Some
spectra still show some residual variations even after subtracting off
a double sinusoidal baseline. Nearest neighbor panels overlap by
8$''$, i.e. about half of the APEX beam width at this frequency, so
that next nearest neighbor panels are roughly independent. The CO
source associated with Hubble{\rm V} is clearly resolved in this map
and is extended towards the north-east.
\label{HV}}
\end{figure}

We made a preliminary assessment of the physical conditions in the CO
cloud associated with Hubble{\rm V} assuming local thermodynamical
equilibrium (LTE). In that case, there is one excitation temperature,
$T_{\rm ex}$ responsible for populating the energy levels of the
$^{12}{\rm CO}$ and $^{13}$CO isotopomers. This need of course not be
the case in reality, with the higher-$J$ lines not being thermalized
due to their larger Einstein $A$ coefficients. In the following, we
will assume the isotopic ratio $X={^{12}{\rm CO}}/{^{13}{\rm CO}}=60$
\citep{lp93,saz02}, in which case the optical depths of $^{12}{\rm
CO}$, denoted by $\tau_{12}$, and $^{13}{\rm CO}$, denoted by
$\tau_{13}$, obey the relation $\tau_{12} = X\,\tau_{13}$. The
calculated line intensities are coupled to the observed quantities by
the beam filling factor $f_{\rm b}$, which we assume to be the same
for the $^{12}{\rm CO}$ and $^{13}{\rm CO}$ emission. Furthermore, we
will assume that both the $^{12}{\rm CO}$ and the $^{13}{\rm CO}$
emission arises in the same region so that $\Omega_{\rm source}$, the
solid angle spanned on the sky by the CO source, is the same for both
isotopomers. This is to keep this preliminary modeling as simple as
possible since there is no physical reason why $f_{\rm b}$ and
$\Omega_{\rm source}$ should be the same for all transitions. We can
use this source solid angle to correct the observed emission line
brightness temperatures for beam dilution using the relation $T'_{\rm
mb} = T_{\rm mb} (\Omega_{\rm source} + \Omega_{\rm beam})/\Omega_{\rm
source}$, with $\Omega_{\rm beam}$ the beam solid angle. The main-beam
brightness temperature of an observed transition can be written as
\begin{equation}
T'_{{\rm mb},i} = f_{\rm b}\,\left( 1 - e^{-\tau_i} \right) \frac{h
\nu_i}{k} \left( \frac{1}{e^{h \nu_i/kT_{\rm ex}}-1} - \frac{1}{e^{h
\nu_i/kT_{\rm cmb}}-1} \right),
\end{equation}
with $\nu_i$ the frequency of the transition, $\tau_i$ its optical
depth, and $T_{\rm cmb} = 2.725$~K the background radiation
temperature.

\begin{table}
\caption{Comparison of the LTE model with the measured intensities of
the $^{12}$CO and $^{13}$CO transitions. All intensities have been
corrected for beam dilution using $\Omega_{\rm source} = 209\,{\rm
arcsec}^2$.
\label{tablte}}
\begin{center}
\begin{tabular}{|ccc|} \hline
transition & measurement (K) & model (K) \\ \cline{1-3}
$^{12}$CO(1$\rightarrow$0) & $1.72 \pm 0.11$ & 1.73 \\
$^{12}$CO(2$\rightarrow$1) & $1.69 \pm 0.08$ & 1.66 \\
$^{12}$CO(3$\rightarrow$2) & $1.56 \pm 0.05$ & 1.57 \\
$^{12}$CO(4$\rightarrow$3) & $1.50 \pm 0.09$ & 1.48 \\
$^{13}$CO(1$\rightarrow$0) & $0.11 \pm 0.07$ & 0.11 \\ \cline{1-3}
\end{tabular}
\end{center}
\end{table}

\begin{figure}
\begin{center}
  \includegraphics[width=7cm,angle=0]{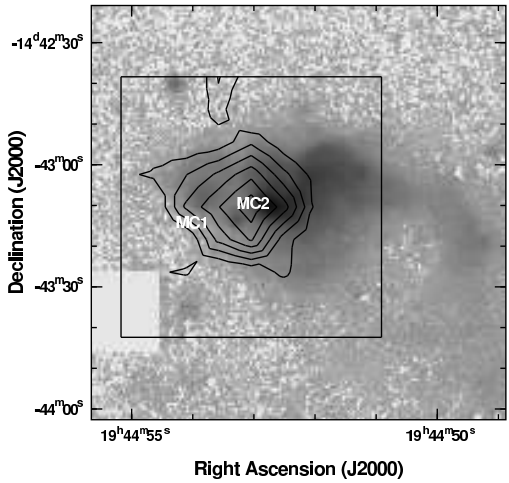}
\caption{Spatial distribution of the CO(3$\rightarrow$2) emission of
Hubble\rm{V} at 18$''$ resolution (contours) plotted over an H$\alpha$
image (greyscale). The square indicates the region mapped by our APEX
observations. The contour values are 1 (which corresponds to
3$\sigma$), 1.5, 2, 2.5, 3.0, and 3.5~K~km~s$^{-1}$. The
CO emission is not centered on the H$\alpha$
emission. The positions of the CO clouds MC1 and MC2, identified by
\cite{w94}, are indicated in the figure.
\label{spatial}}
\end{center}
\end{figure}
Using a non-linear minimisation routine, we simultaneously fitted a
Gaussian model for the spatial distribution of the CO emission of
Hubble{\rm V}, convolved with the APEX beam, which constrains the
source solid angle $\Omega_{\rm source}$ and the parameters $T_{\rm
ex}$, $f_{\rm b}$, and $\tau_{12}$ to the $^{12}$CO(3$\rightarrow$2)
map, presented in fig. \ref{HV}, and to the $^{12}$CO(3$\rightarrow$2)
brightness temperature measured by us, and the
$^{12}$CO(1$\rightarrow$0), $^{12}$CO(2$\rightarrow$1),
$^{12}$CO(4$\rightarrow$3), and $^{13}$CO(1$\rightarrow$0) brightness
temperatures presented in \cite{i03}.  We then used the best fitting
values for $\Omega_{\rm source}$, $T_{\rm ex}$, $f_{\rm b}$, and
$\tau_{12}$ to generate 10000 mock data sets with added Gaussian noise
on each of the observed quantities, using the measured 1$\sigma$
uncertainties on the measured quantities as estimates for the
dispersions of each of the noise distributions. These mock data sets
were analysed the same way as the original set, allowing us to
estimate the 1$\sigma$ errors on the derived quantities. This way, we
find that the parameter values $\Omega_{\rm source} = 209 \pm 50\,{\rm
arcsec}^2$, $T_{\rm ex} = 49 \pm 27\,{\rm K}$, $\tau_{12} = 3.7 \pm
2.3$, and $f_{\rm b}= 0.04 \pm 0.03$ provide the best fit to the whole
data-set. There is a large degree of degeneracy between the parameters
of this model, e.g. between $T_{\rm ex}$ and $f_{\rm b}$. This is
reflected by the very large errorbars on these quantities. Still, the
minimisation routine converges to the same solution independent of the
starting point of the minimisation which proves that the minimum of
the $\chi^2$ is well defined. Moreover, this temperature estimate
agrees with the dust temperature derived from the ratio of the $60
\mu$ and $100 \mu$ IRAS flux densities, $f_\nu(60)=7.89$~Jy and
$f_\nu(100)=11.81$~Jy, of Hubble{\rm V}, $T_{\rm dust} \approx
40$~K. This estimate was derived assuming a single temperature
component and a $\lambda^{-1}$ emissivity law. Given the apparently
rather high temperature of this CO emission cloud, observations of
higher-$J$ transitions, e.g. with FLASH or the future SIS heterodyne
receivers, are required for a more precise assessment of its physical
properties using more sophisticated LVG models, taking into account
non-LTE effects. Also, some of the published high-$J$ transition
temperatures, such as $^{12}$CO(4$\rightarrow$3) value of \cite{i03},
may be affected by the small area that was mapped. If some of the
emission was missed, this may lead to an underestimation of the
brightness temperature of these transitions.

Using the $^{12}$CO(3$\rightarrow$2) FWHM line-width, $\Delta V$, in
km~s$^{-1}$, and the radius of the emission region, $R$, in parsec, we
can also estimate the virial mass of Hubble~V as $M_{\rm vir} \approx
190 R (\Delta V)^2$, in solar masses \citep{m88}. For $\Delta V =
6.0$~km~s$^{-1}$, $R = \sqrt{\Omega_{\rm source}}/2.355 =6.1'' =
13.9$~pc, we find $M_{\rm vir} \approx 9.5 \times 10^4\,M_\odot$.
Using the relation $M_{\rm dust} = 1.27 f_\nu(100) D^2
(\exp(144\,K/T)-1)\,M_\odot$ \citep{blg02} for the dust mass, with $D$
the distance in Mpc, we find $M_{\rm dust} \approx
130\,M_\odot$. Using the metallicity-dependent gas versus dust mass
relation (eqn. (6) in \cite{blg02}), this yields $M_{\rm gas} \approx
9 \times 10^4\,M_\odot$, in good agreement with the virial mass. This
is much less than the estimate of the total gas mass $M_{\rm gas} = 10
\pm 5 \times 10^5\,M_\odot$ of \cite{i03}. This is most likely because
we are measuring quantities (radius, velocity dispersion) that pertain
only to the CO emission region : any mass distribution outside this
region, which was taken into account by \cite{i03}, would not have a
very large effect on the virial mass estimate. Hubble~V is also very
near the line-width versus size relation traced by the Milky Way and
LMC molecular clouds \citep{h95,m90}. We estimate the
$^{12}$CO(3$\rightarrow$2) to H$_2$ conversion at $5.4-5.8 \times
10^{20}$ cm$^{-2}$ (K\,km\,s$^{-1}$)$^{-1}$, depending on whether the
deconvolved or the observed brightness is used. 

\subsection{Comparison with previous observations}

In Figures~\ref{HIV} and \ref{HV}, we plot the (unscaled) HI spectra
derived by \citet{deblok06}, of the same regions as our observations
on top of the $^{12}$CO(3$\rightarrow$2) spectra. One can see that
apart from S28, both emission lines are coincidental although the HI
emission is systematically broader than the $^{12}$CO(3$\rightarrow$2)
emission. In Figure~\ref{spatial}, we plot the spatial distribution of
the $^{12}$CO(3$\rightarrow$2) emission of Hubble\rm{V} at 18$''$
resolution. We find a different morphology than the one derived by
\citet{i03}, however it is in accordance with their [C\rm{II}]
emission, found in the same paper. A possible cause might be the
higher system temperature of 2460K during their observation, producing
more noise which might cause a shift in the spatial distribution. The
main emission peak in our map corresponds to the molecular cloud MC2,
first detected by \cite{w94}; the eastward extension 
corresponds only very roughly to MC1. 

With our observations we prove that APEX is very suited for deriving
spatially extended, high signal-to-noise maps of emission-line regions
in Local Group dwarf galaxies, where one can achieve a spatial
resolution of a few tens of parsecs.

\begin{acknowledgements}
We thank Erwin de Blok for kindly permitting us to use the H$\alpha$
image of Hubble~V and the radio spectra of the different NGC6822
fields. We thank the referee, Jonathan Braine, for his valuable
remarks.
\end{acknowledgements}

\end{document}